\documentstyle[12pt]{article}

\begin{document}

\newcommand{\bphi}{\Phi }
\newcommand{\bpsi}{\Psi }
\def\'#1{\if#1i{\accent19\i}\else{\accent19#1}\fi}

\newcommand{\APSREF}{1}
\newcommand{\LETTER}{0}

\def\daga#1{{#1\mkern -9.0mu /}}
\newcommand{\Eq}[1]{Eq.~(\ref{#1})}
\newcommand{\Ref}[1]{Ref.~\cite{#1}}
\newcommand{\beq}[1]{\begin{equation}\label{#1}}
\newcommand{\eeq}{\end{equation}}
\newcommand{\bdm}{\begin{displaymath}}
\newcommand{\edm}{\end{displaymath}}
\newcommand{\beqa}[1]{\begin{eqnarray}\label{#1}}
\newcommand{\eeqa}{\end{eqnarray}}
\newcommand{\bdma}{\begin{eqnarray*}}
\newcommand{\edma}{\end{eqnarray*}}

\ifnum\APSREF=1
\newcommand{\prd}[3]{Phys. Rev. D{\bf #1}, #2 (#3)}
\newcommand{\prb}[3]{Phys. Rev. B{\bf #1}, #2 (#3)}
\newcommand{\physrep}[3]{Phys. Rep. {\bf #1}, #2 (#3)}
\newcommand{\plb}[3]{Phys. Lett. {\bf B#1}, #2 (#3)}
\newcommand{\npb}[3]{Nucl. Phys. {\bf B#1}, #2 (#3)}
\newcommand{\np}[3]{Nucl. Phys. {\bf #1}, #2 (#3)}
\newcommand{\prl}[3]{Phys. Rev. Lett. {\bf #1}, #2 (#3)}
\newcommand{\rmp}[3]{Rev. Mod. Phys. {\bf #1}, #2 (#3)}
\newcommand{\ibid}[3]{{\em ibid.} {\bf #1}, #2 (#3)}
\newcommand{\astropj}[3]{Ap. J. {\bf #1}, #2 (#3)}
\newcommand{\annphys}[3]{Ann. Phys. {\bf #1}, #2 (#3)}
\newcommand{\repprp}[3]{Rep. Prog. Phys. {\bf #1}, #2 (#3)}
\newcommand{\jmp}[3]{J. Math. Phys.  {\bf #1}, #2 (#3)}
\newcommand{\zphysc}[3]{Z. Phys. C  {\bf #1}, #2 (#3)}
\newcommand{\ijmpa}[3]{Int. J. Mod. Phys. A  {\bf #1}, #2 (#3)}
\newcommand{\mpla}[3]{Mod. Phys. Lett. A{\bf #1}, #2 (#3)}
\newcommand{\jmpa}[3]{J.  Phys A: Math. Gen.  {\bf #1}, #2 (#3)}
\newcommand{\commp}[3]{Commun. math. Phys. {\bf #1}, #2 (#3)}

\else
\newcommand{\prd}[3]{Phys. Rev. D{\bf #1} (#3) #2 }
\newcommand{\prb}[3]{Phys. Rev. B{\bf #1} (#3) #2 }
\newcommand{\physrep}[3]{Phys. Rep. {\bf #1} (#3) #2 }
\newcommand{\plb}[3]{Phys. Lett. {\bf B#1} (#3) #2 }
\newcommand{\npb}[3]{Nucl. Phys. {\bf B#1} (#3) #2 }
\newcommand{\np}[3]{Nucl. Phys. {\bf #1} (#3) #2 }
\newcommand{\prl}[3]{Phys. Rev. Lett. {\bf #1} (#3) #2 }
\newcommand{\rmp}[3]{Rev. Mod. Phys. {\bf #1} (#3) #2 }
\newcommand{\ibid}[3]{{\em ibid.} {\bf #1} (#3) #2 }
\newcommand{\astropj}[3]{Ap. J. {\bf #1} (#3) #2 }
\newcommand{\annphys}[3]{Ann. Phys. {\bf #1} (#3) #2 }
\newcommand{\repprp}[3]{Rep. Prog. Phys.  {\bf #1} (#3) #2 }
\newcommand{\jmp}[3]{J. Math. Phys.   {\bf #1} (#3) #2 }
\newcommand{\zphysc}[3]{Z. Phys. C   {\bf #1} (#3) #2 }
\newcommand{\ijmpa}[3]{Int. J. Mod. Phys. A  {\bf #1} (#3) #2 }
\newcommand{\mpla}[3]{Mod. Phys. Lett. A{\bf #1} (#3) #2 }
\newcommand{\jmpa}[3]{J.  Phys A: Math. Gen.   {\bf #1} (#3) #2 }
\newcommand{\commp}[3]{ Commun. math. Phys.  {\bf #1} (#3) #2 }
\fi

\ifnum\LETTER=0
\newcommand{\newsection}[1]{\section{#1}\setcounter{equation}{0}}
\fi

\newcommand{\aslash}[1]{{\rlap/#1}}
\newcommand{\splash}[1]{{#1\mkern -9.0mu /}}
\newcommand{\splashh}[2]{{#2\mkern -#1mu /}}

\baselineskip=20pt

\begin{flushright}
Preprint IFUNAM:\\
FT95-77  May/95 .\\
hep-th/yymmnn
\end{flushright}

\vskip0.75cm

\begin{center}{\LARGE\bf
 Self-dual non-Abelian  vortices in a
  $ \bphi^2$ Chern-Simons theory.\\ \vskip0.2in
}\end{center}
\begin{center}
by

{\bf{ Armando Antill\'on
$^a$\footnote{e-mail:armando@ce.ifisicam.unam.mx},
 Joaqu\'in Escalona$^b$\footnote{e-mail:
joaquin@llei.ifisicam.unam.mx},
Gabriel Germ\'an$^a$\footnote{e-mail:gabriel@ce.ifisicam.unam.mx},
 \\and
 Manuel Torres$^c$\footnote{e-mail: manuel@teorica1.ifisicacu.unam.mx
}
}}\\\vskip0.2in
{\small\it  $^a$ Laboratorio de Cuernavaca, Instituto de F\'isica,
Universidad Nacional  Aut\'onoma \\
\vspace{-2mm}
 de M\'exico,   Apdo. Postal 48-3,
62251 Cuernavaca, Morelos,   M\'exico.} \\
{\small\it  $^b$ Facultad de Ciencias, Universidad Aut\'onoma del
Estado de
Morelos, \\
\vspace{-2mm}
Apdo. Postal 396-3, 62250 Cuernavaca, Morelos,   M\'exico.}\\
{\small\it  $^c$Instituto de F\'isica,  Universidad Nacional
Aut\'onoma  de
M\'exico, \\
\vspace{-2mm}
Apdo. Postal 20-364, 01000  M\'exico, D.F., M\'exico.}

\vskip0.25in
\noindent
{\bf{Abstract}}
\end{center}
\vskip0.05in
\setlength{\baselineskip}{0.2in}
We study a non-Abelian Chern-Simons gauge  theory in $ 2+ 1$
dimensions  with the inclusion of an  anomalous magnetic interaction.
For a particular relation  between the Chern-Simons (CS)  mass and
the anomalous  magnetic coupling the  equations for the gauge fields
reduce from second- to first  order differential equations of the
pure
CS type.  We derive the Bogomol'nyi-type  or self-dual  equations
for a $\bphi^2$ scalar  potential, when the scalar and topological
masses are equal. The corresponding  vortex solutions carry
magnetic flux that is not quantized  due to the  non-toplogical
nature of the solitons. However,  as a consequence of the
quantization  of the CS term, both the electric charge and
angular momentum are quantized.
\vskip 0.5cm
\begin{center}
\noindent
\rule[.1in]{3.0in}{0.002in}
\end{center}
\vskip1cm
keywords: Chern-Simons,  non-topological, non-abelian,
vortices.
\newpage
\baselineskip=20pt

Gauge field theories present a rich spectrum of finite energy
(or finite  action ) classical solutions; such as  vortices,
monopoles and
instantons. These classical solutions can be classified as
topological or
non-topological; depending of the
origin of the stability mechanism \cite{review1}.
Among these theories, the self-dual theories deserve
special attention.  Self-duality refers to theories in which
the interactions have particular forms and special strengths such
that the equations of motion reduce from  second- to first-order
differential
equations; these configurations  minimize the energy (or the action).
For example the  Abelian-Higgs model admits topological
solitons of the vortex type \cite{nielsen}; furthermore, when the
parameters are chosen to make the vector and scalar masses
equal  the  vortices satisfies a set of Bogomol'nyi-type
or self-duality-type equations \cite{bogo}. This self-dual
vortex solutions have   also been found  for  the non-Abelian Higgs
theory \cite{cuglian}. In the self-dual point  the vortices  become
non-interacting and static multisoliton solutions may be expected
\cite{taubes}.

Recently there has been considerable interest in a new class
of self-dual theories, the self-dual Chern-Simons  (CS) theories
in (2 +1) dimensions,
which involve charged scalar fields minimally coupled to
massive gauge fields whose dynamics is solely
provided by the CS term, instead of the Maxwell term
  \cite{review2}.
These theories,  where
the kinetic action for the gauge field is solely
provided by  the Chern-Simons
term is known as the pure CS theory  (PCS) \cite{deser,review3}.
An interesting feature of these self-dual  theories is that they
permit a realization with either relativistic or nonrelativistic
dynamics; furthermore the presence of the CS  term produce
interesting effects: the magnetic vortices acquire electric charge
and fractional spin \cite{paul1}.
In the case of the relativistic theory,
self-dual  vortex solutions have been found for a  sixth order
 Higgs potential  in  both Abelian  \cite{hong,jackwein}
and non-Abelian \cite{cuglian2} theories.

One can pose the question whether there exist  self-dual
models  in which the gauge field Lagrangian includes
both the Maxwell and the CS term.  Self-dual vortex solutions
can be constructed in such
 a Maxwell-Chern-Simons  gauge theory  if one adds
 a magnetic moment interaction between the
scalar  and the gauge fields \cite{torres}.
It was shown that for a special relation between the   CS mass and
the anomalous magnetic coupling, the equations for the gauge fields
reduce from second- to first-order differential equations,
similar to those of the pure CS theory. Furthermore,
it was demonstrated that non-topological  charged
vortices satisfy a set of
Bogomol'nyi-type or self-duality  equations for a quadratic potential
$V( \bphi) = (m^2/2) \bphi^2$, when  $m$ and the topological masses
are equal \cite{torres2}. This model possess a local $U(1)$ symmetry,
so we
will refer to it as the
Abelian  $ \bphi^2$ model.

In this work we present an extension of the  $ \bphi^2$ model
 to the non-Abelian case, finding first-order self-dual equations
which can  be  seen to admit  vortex  solutions carrying
magnetic flux, electric charge and spin.
The same as in the Abelian model   self-duality
is attained for a quadratic scalar potential, consequently the
solitons are non-topological; so  the magnetic
flux is not quantized. However, it is interesting to note that
as a consequence  of the quantization of the CS coefficient
$\kappa$ in the non-Abelian case \cite{deser}, both the electric
charge and the spin become quantized.

We shall consider  for simplicity  a  non-Abelian gauge theory in $2
+1 $
dimensions
with $SU(2)$ symmetry, though,
the arguments that follow can be easily generalized  to   an  $SU(N)$
theory.
Thus, we have a   theory of gauge fields ${\bf A}_\mu$  coupled to
two scalar
field
$\bphi$ and $\bpsi$  in the adjoint representation.
 The   Lagrangian  describing our model reads

\begin{eqnarray}\label{lagrangian}
  {\cal L}  =
  & {1 \over 2}&  D_\mu    \bphi \cdot D^ \mu  \bphi  \, + \,
 {1 \over 2} D_\mu   \bpsi \cdot D^ \mu   \bpsi
- V(  \bphi , \bpsi )  \cr
 & - &  {1 \over 4}  {\bf  G} _{\mu \nu}  \cdot  {\bf G}^{\mu \nu} \,
+
   {\kappa \over 4} \epsilon^{\mu \nu \alpha}
\left(  {\bf G}_{\mu\nu } \cdot {\bf A}_\alpha  - {e \over 3}
 {\bf A}_\mu \cdot \left(  {\bf A}_\nu \wedge  {\bf A}_\alpha
\right)\right)
    \,  , \end{eqnarray}

where the Minkowski-space metric is
$g_{\mu \nu}  = diag \, (+1,-1,-1); \, \mu  =(0,1,2),  \,$
with

\begin{equation}\label{gmunu}
 {\bf G}_{\mu \nu} = \partial_\mu {\bf A}_\nu - \partial_\nu {\bf
A}_\mu
+ e {\bf A}_\mu\wedge {\bf A}_\nu
\, .
\end{equation}

In  all these equations $\bphi$,  $\bpsi$, and ${\bf A}_\mu$ denote
triplets  in isospin-space;  $e.g.$
 $\bphi = (\phi^1,\phi^2\phi^3)$,
 $ {\bf A}_\mu = (A_\mu^1,A_\mu^2,A_\mu^3)$, etc.
The covariant derivative  $D_\mu$ includes both the usual minimal
coupling plus the anomalous magnetic contribution:

\begin{equation}\label{dercov}
 D_\mu   \bphi = \big( \partial_\mu  \bphi  +  e {\bf A}_\mu \wedge
\bphi  +
 {g \over 4 } \epsilon_{\mu\nu\alpha} {\bf G}^{\nu\alpha} \wedge
\bphi
   \big)   \, ,
\end{equation}

with $g$ the anomalous  magnetic moment.
It has been previously  proved,   that  for Abelian  gauge theories
 the magnetic moment interaction  can be
incorporated into  the covariant derivative,  even for spinless
particles
\cite{stern,paul2,torres}.
\footnote{The magnetic moment interaction
term in CS  gauge theories has also been considered
in some other context, see references \cite{otros}.}
Here  \Eq{dercov} shows that in the non-Abelian case  the covariant
coupling to the scalar field can also be modified by the inclusion of
an anomalous magnetic term.
This extra term is  consistent with the Lorentz and gauge
covariance of the theory, but give rise to the
breaking of $P$ and $T$ symmetries.

We choose for the potential

\begin{equation}\label{potential}
 V(  \bphi , \bpsi )  = V_1(\bphi) + V_2(\bpsi) + \lambda  \left(
\bphi \cdot
\bpsi \right)^2
\, ,
\end{equation}
where the last term is  selected  in such a  way that   the energy is
minimized
by configurations in which  $\bphi$ and $\bpsi$ are
orthogonal
in isospin space.   We  also require $V_2(\bpsi)$ to be  minimized at
a
non-zero value
$\bpsi^2 = v^2 $, $i.e.$  $V_2(|\bpsi| = v) = 0$.
For the moment we leave $V_1(\bphi)$ free; we  shall see that it is
determined
by the
Bogomol'nyi equation to be    $V_1(\bphi) = (\kappa/2) \, \bphi^2 $.

The field Lagrangian in \Eq{lagrangian} leads to gauge
covariant equations of motion.  However,  the Lagrangian itself
is not   gauge invariant.
Indeed, as a response  of  a gauge transformation $U$,
the change in the  action $S$  is
$S \rightarrow S + \kappa ( 8 \pi^2 /e^2) \omega \,  (U)$, where
$\omega (U)$
is the winding number of the gauge  transformation. The
corresponding quantum theory is well defined if the change
in the action is a multiple of $2 \pi$; this leads to the
quantization
condition of the topological  mass \cite{deser}:

\begin{equation}\label{cuank}
 \kappa  \, =  \, {e^2 \over 4 \pi} \, n \, ,
\qquad n \in {\bf Z} \, .
\end{equation}

The equations of motion for the Lagrangian in  \Eq{lagrangian} are

\begin{equation}\label{eqsmot1}
D_\mu D^\mu   \bphi  =  - {\delta V \over \delta \bphi} \, ,  \qquad
D_\mu D^\mu   \bpsi  =  - {\delta V \over \delta \bpsi}
 \,  ,
\end{equation}

\begin{equation}\label{eqsmot2}
 \epsilon_{\mu \nu \alpha}  \nabla^\mu \big [ {\bf G}^\alpha +
{ g \over 2 e}  {\bf J}^\alpha \big]
     =  {\bf J} _\nu - \kappa {\bf  G}_\nu
      \,  .
\end{equation}

The last equation has been written in terms of the dual field,
$ {\bf G}_\mu \equiv {1 \over 2} \epsilon_{\mu \nu \alpha}  {\bf
G}^{\nu
\alpha}$.
We notice that in the previous equation $\nabla^\mu$ includes only
the
contribution
on the gauge potential (compare with the full covariant derivative
\Eq{dercov}
):

\begin{equation}\label{dernabla}
 \nabla_\mu   \bphi = \left( \partial_\mu  \bphi  +  e {\bf A}_\mu
\wedge
\bphi
   \right)   \, ,
\end{equation}

 and the conserved matter current  is given by

\begin{equation}\label{current}
 {\bf J}_\mu =  e \left( D_\mu \bphi \wedge \bphi + D_\mu \bpsi
\wedge \bpsi
\right)
   \,  .
\end{equation}

In the case of the Abelian  theory the  gauge field equations reduce
from second- to first-order differential equations
\cite{stern,torres},
similar to those of the PCS type \cite{review3} when
the relation   $\kappa = - {2 e \over g}$ holds. The same is true in
the
non-Abelian theory; indeed
 we   notice that if the   relation

\begin{equation}\label{kappag}
 \kappa = - {2 e \over g}
\end{equation}

 holds, then it is clear that the  \Eq{eqsmot2}  is  solved
identically if we
choose
the first order ansatz

\begin{equation}\label{eqcs}
 {\bf G}_\mu  = {1 \over \kappa}  {\bf J}_\mu
  \,    ,
\end{equation}

that  has the same structure as   the equations of the PCS theory
\cite{review3}.
We will refer to the previous conditions as the PCS limit. However,
we should
notice that
  the explicit expression
for ${\bf J}_\mu$ differs from the usual expression of the PCS
theory, because
according  to Eqs.  (\ref{current}) and
(\ref{dercov})  ${\bf J}_\mu$ receives contributions from the
anomalous
magnetic moment.
This  PCS
equations (\Eq{eqcs})  imply that any object carrying magnetic flux
($
\Phi_B$) must
also carry electric charge ($Q$), with the two quantities related
as

\begin{equation}\label{fundamental}
 Q =  \kappa   \Phi_B \, .
\end{equation}

In what follows we shall work in the  limit in which
Eqs. (\ref{eqcs}) and  (\ref{kappag}) are  valid,
so we consider  \Eq{eqcs} as the equation of motion for the gauge
fields,
instead of  \Eq{eqsmot2}.

In the  so called  Bogolmol'nyi limit    all the equations of motion
are known
to become first
order differential equations \cite{bogo};  furthermore,  it is
possible  to
write
the equations of motion as self-duality  equations. The existence of
this limit
usually
requires   a specific form  for the scalar potential. To look for the
Bogomol'nyi equations we start from the energy-momentum tensor that
is
obtained by varying the curved-space form  of the action with respect
to the
metric;
this yields

\begin{eqnarray}\label{tmunu}
   T_{\mu \nu} & =&
\left[ \delta_{ab}  - {g^2 \bphi^2 \over  4}
 \left( \delta_{ab} - \hat{\phi}_a  \hat{\phi}_b \right)
 - {g^2 \bpsi^2 \over  4}
 \left( \delta_{ab} - \hat{\psi}_a  \hat{\psi}_b \right)
 \right]
  \left(  G_\mu ^a   G_\nu^b  -  {1 \over 2} g_{\mu \nu}  G_\alpha ^a
G^\alpha_b \right)
      \cr
      \cr
    & +  &    \nabla_\mu \bphi \cdot   \nabla_\nu   \bphi
-  {1 \over 2}g_{\mu \nu}   \left( \nabla_\lambda   \bphi \right)^2
+ \nabla_\mu \bpsi \cdot   \nabla_\nu   \bpsi
-  {1 \over 2}g_{\mu \nu}   \left( \nabla_\lambda   \bpsi \right)^2
 + g_{\mu\nu}   V (\bphi,\bpsi) \, , \cr
 & &     \end{eqnarray}
where  $\hat{\phi}_a = \phi_a/ |\bphi|$ ,
 $\hat{\psi}_a = \psi_a/ |\bpsi|$
 and we recall that
$\nabla_\mu $ includes only the gauge
potential contribution (\Eq{dernabla}); we reserve the letters
$a,b,c,...$ for
isospin
indices and $i,j,k,...$ for space indices.
Notice  that both  the Chern-Simons  and linear terms in $g$
do not appear explicitly in $T_{\mu\nu}$. This is a consequence of
the fact
that  these
terms  do not make use of the space-time  metric tensor $g_{\mu\nu}$;
then ,
when  $g_{\mu\nu}$ is varied  to produce $T_{\mu\nu}$  no
contributions  arise
from
these terms \cite{review3}.

As it is well-known,  finite energy  determines the   asymptotic
behavior
of  the fields at spatial  infinity.  Thus, we demand that every term
in \Eq{tmunu} vanish as $\rho \rightarrow \infty$:

\begin{eqnarray}\label{cond1}
\lim_{\rho\rightarrow \infty}{ \nabla_\mu \bphi } &= &0  \, , \qquad
\lim_{\rho\rightarrow \infty}{ \nabla_\mu \bpsi = 0}  \, , \cr
\lim_{\rho\rightarrow \infty}{{\bf G}_{\mu}} & =&0  \, ,
\qquad \lim_{\rho\rightarrow \infty} V (\bphi,\bpsi)= 0
\, .
\end{eqnarray}

The  equations of motion and the energy-momentum tensor
are too complicated
in their actual form.  So, in order to find static finite-energy
solutions we
shall assume some simplifying properties
for the fields; later, an ansatz consistent with these conditions
will be presented. We look for conditions in such a way that the
field $\bpsi$   plays no dynamical role, but it points along
a constant direction in isospin-space that can be used to define a
gauge
invariant Abelian  field strength.
Thus, we take the field $\bpsi$ as a constant
everywhere in space; we expect that any other configuration with a
non-constant
  $\bpsi$  should  lead to  a greater
energy.
 Furthermore, we also assume that the field $\bpsi$ is parallel in
isospin
space to the gauge fields.
We shall then  look for the self-duality equations under the
following
conditions (valid everywhere):

\begin{eqnarray}\label{cond2}
 { D_\mu \bpsi } &= &0  \, , \qquad
 \bpsi^2 = v^2  \, , \cr
 \bpsi \cdot \bphi & =&0  \, , \qquad
\bpsi  \, || \,  {\bf A}_\mu  \, .
\end{eqnarray}
Notice that under these conditions the last two terms of the
potential
(\Eq{potential})
vanish identically and  that
the field $\bpsi$  does not contribute to the total
energy. Using the  previous conditions and Eqs.  (\ref{current})  and
(\ref{eqcs}) one can  also see that    $\bphi \cdot {\bf G}_\mu = 0$.
Due to the fact that  $\bphi$  and   ${\bf G}_\mu $  are mutually
orthogonal we
can not project
${\bf G}_\mu $  along the direction of   $\bphi$  in order to define
the
physically
observable electromagnetic field $F_{\mu\nu}$. This is the reason why
we
introduced  a second
scalar field $\bpsi$; hence we propose to construct the gauge
invariant  Abelian field  strength by setting

\begin{equation}\label{fmunu}
 F_{\mu\nu} = \hat{\bpsi} \cdot {\bf G}_{\mu\nu} \, ,  \qquad
\hat{\bpsi}  =  {\bpsi \over |\bpsi|}
\,  .
\end{equation}

The energy  density   ($T_{00}$) for a static configuration can be
written
in terms of the ``cromo-electric''   $E^a _i=  - \epsilon_{ij} G^a_j
$ and
 ``cromo-magnetic''  $B^a = G^a_0$ fields as

\begin{eqnarray}\label{t00}
   T_{00}     =     {1 \over 2}  \left( 1 - {e^2 \over \kappa^2 }
\bphi^2
\right)
 \left( {\bf E}_i^2   + {\bf B}^2 \right)
    +      {1 \over 2}  \left( {\bf A}_0 \wedge \bphi \right)^2 +
    {1 \over 2}  ( \nabla_i  \bphi  )^2    +  V_1 (\bphi)
   \,.    \end{eqnarray}
Defining a reduced gauge potential

\begin{equation}\label{reda}
\bar {{\bf A}}_\mu  =  {\bf A}_\mu   + { 1 \over e} \hat{\bphi }
\wedge
\partial_\mu
\hat{\bphi}
\, ,
\end{equation}
with $ \hat{\bphi} = \bphi/|\bphi|$,
we can combine  Eqs.  (\ref{eqcs}) and  (\ref{current}) to express
${\bf
G}_\mu $
in terms of  $ \bar {{\bf A}}_\mu $

\begin{equation}\label{relga}
G^a_\mu = -   \kappa \bphi^2  C
 \left( \delta_{ab} - \hat{\phi_a}  \hat{\phi_b}  \right)
\bar{A}_\mu^b
\, ,
\end{equation}
where  $C =   e^2/ \left( \kappa^2 - e^2 \bphi^2 \right) $.
Notice,  that  \Eq{relga} is in agreement with the condition
  $ \bphi \cdot {\bf G}_\mu = 0$,
 therefore  $ \bphi \cdot {\bf B} = 0$ and  $ \bphi \cdot {\bf E}_i =
0$.

We can now use the $\mu = 0$ component of  the previous equation  to
eliminate
${\bf A}_0$ from \Eq{t00}.    Hence, the part of $T_{00}$ depending
on ${\bf
B}$ and
${\bf A}_0$ can be recast as

\begin{equation}\label{rel1}
 {1 \over 2}  \left( 1 - {e^2 \over  \kappa^2}  \bphi^2 \right)  {\bf
B}^2
+  {1 \over 2}  \left( {\bf A}_0 \wedge \bphi \right)^2
\, = \, {1  \over  2 C \bphi^2 }  {\bf B}^2
\, ,
\end{equation}
where we  have  used the condition  $ \bphi \cdot {\bf B} = 0$.

Likewise,  using the $\mu = i$ components of \Eq{relga}
 we can  combine the  parts of $T_{00}$ involving the electric field
and
$\nabla_i \bphi$
into the following form

\begin{equation}\label{rel2}
 {1 \over 2}  \left( 1 - {e^2 \over \kappa^2 }  \bphi^2 \right)
\left( E_i^a \right)^2
+     {1 \over 2}  ( \nabla_i  \bphi  )^2  =
{1 \over 2 \bphi^2 } \left( \bphi \cdot \partial_i \bphi \right)^2
+ {1 \over 2} {\kappa^2  \bphi^2    C }
\left( \delta_{ab} - \hat{\phi}_a  \hat{\phi}_b \right)
\bar{A}^a_i \bar{A}^b_i
\, .
\end{equation}

Substituting  the results of  Eqs. (\ref{rel1}) and    (\ref{rel2})
into
\Eq{t00}
 we can write down   the energy $E = \int d^2x T_{00}$ as

\begin{equation}\label{ener1}
E  = \int d^2x \left( { 1\over 2 C  \bphi^2} {\bf B}^2
+ {1 \over 2} \left[ {1 \over \bphi^2}\left( \bphi \cdot \partial_i
\bphi
\right)^2  +
 { \kappa^2   \bphi^2     C}
\left( \delta_{ab} - \hat{\phi}_a \hat{\phi}_b \right) \bar{A}^a_i
 \bar{A}^b_i  \right]  + V_1(\bphi)
\right)
\, .
\end{equation}

The energy written in this form is similar to the expression that
appears in
the
Nielsen-Olesen model. Thus,   starting from \Eq{ener1} we  can
follow  the
usual
 Bogomol'nyi-type arguments in order to    obtain the self-dual
limit.  The
energy may then
be rewritten, after an integration by parts, as

\begin{eqnarray}\label{ener2}
   E      &= &   {1\over 2}   \int d^2x \left[
 {1 \over C \bphi^2 }  \left( B_a  \mp   \kappa \sqrt{C} \bphi^2
\hat{\psi}_a
\right)^2
+  |\psi_a {\partial_{\pm} \bphi^2 \over 2 |\bphi| }  - i \kappa
|\bphi|
\sqrt{C}
\left( \delta_{ab} - \hat{\phi}_a \hat{\phi}_b \right)
\bar{A}^b_{\pm} |^2
\right]
            \cr
             \cr
            &+ & \int d^2 x  \left[ V_1(\bphi) - {1 \over 2} \kappa^2
\bphi^2
\right]
\, \pm  \int d^2 x \psi_a \left[  {\kappa \over \sqrt{C}} B_a  -
{\kappa
\sqrt{C} \over 2}
\epsilon_{ij} \left( \partial_i \bphi^2 \right) \bar{A}^a_j
\right]  \, ,
\end{eqnarray}
where   $\partial_{\pm} = \partial_1 \pm i \partial_2$,
 $\bar{\bf{A}}_{\pm} =  \bar{\bf{A}}_1 \pm i  \bar{\bf{A}}_2$ and
 conditions (\ref{cond2}) have been  repeatedly used.
Following \cite{cuglian} we choose a gauge in  such a way that the
conditions

\begin{equation}\label{gauge}
\epsilon_{ij} \hat{\bpsi} \cdot \left(  {\bf A}_i  \wedge {\bf A}_j
\right) = 0
\, ,
\qquad
\epsilon_{ij} \hat{\bpsi} \cdot \left(  \partial_i \hat{\bphi}
\wedge
\partial_j \hat{\bphi} \right) = 0
\end{equation}
 hold. Hence,   recalling the definition of the Abelian
field strength \Eq{fmunu},
one has that  the
 last two terms  in \Eq{ener2}  can be related  to the magnetic flux
of the
vortex configuration

\newpage

\begin{eqnarray}\label{ener3}
 & & \! \! \!\! \! \! \pm  \int d^2 x \psi_a \left[  {\kappa \over
\sqrt{C}}
B_a  - {\kappa \sqrt{C} \over 2}
\epsilon_{ij} \left( \partial_i \bphi^2 \right) \bar{A}^a_j
\right]   =  \pm  \kappa  \int d^2 x  \psi^a \epsilon_{ij} \partial_i
\left[ {1 \over \sqrt{C}}  A_j^a + \Lambda_j^a \right] =
\cr \cr
& & \! \! \! \!\! \! \!\!\!   \pm  \kappa  \oint_{r = \infty}
\bpsi \cdot \left(  {1 \over \sqrt{C}}  {\bf A}_i  +  {\bf \Lambda}_i
\right)
dl_i    =
 \pm {\kappa^2 \over e} \oint_{r = \infty}
\bpsi \cdot  {\bf A}_i dl_i  =  \pm {\kappa^2 \over e} \int d^2x
\bpsi \cdot
{\bf B}
 \equiv {\kappa^2 \over e} |\Phi_B|
\, ,   \cr
& &
\end{eqnarray}
where we  defined
$\Lambda^a_i  = - e^2 \left[  \kappa - \sqrt{\kappa^2 - e^2 \bphi^2
}\right]
\epsilon_{abc} \hat{\phi}_b  \partial_i \hat{\phi}_c    $
   and   we  have  used    the gauge conditions  (\ref{gauge})
and the fact that for
 any non-topological soliton the asymptotic conditions are such that
$ \bphi
\to 0$ at spatial
infinity. Thus,  along the  line integral:  $\Lambda_i^a$ vanishes
and $1/\sqrt{C} = \sqrt{ \kappa^2 - e^2 \bphi^2 }/ e \to \kappa/e$.

  We then see from   Eqs. (\ref{ener2})    and  (\ref{ener3})
  that the energy is bounded below; for a fixed value of the magnetic
flux, the
lower  bound is given by   $E \, \geq{\kappa^2 \over e}   \Phi_B$
provided
that the potential  $V_1(\bphi)$
is chosen   as  $V_1(\bphi) =  {m^2 \over 2} \bphi^2$ with the
critical value
$m = \kappa$, $i.e.$ when the scalar and
the the topological masses are equal. Therefore,  in this limit  we
are
necessarily in the
symmetric phase of the theory. From \Eq{ener2} we  see that the lower
bound for
the energy

\begin{equation}\label{ener4}
 E \, = \, {\kappa^2 \over e} | \Phi_B| \, = \, {\kappa \over e} |Q|
  \,  ,
\end{equation}

is saturated when the following self-duality equations are satisfied:

\begin{equation}\label{sd1}
   B_a  \,   =  \, \pm {\kappa e \bphi^2 \over \left[ \kappa^2  - e^2
\bphi^2
\right]^{1/2}}
\,  \hat{\psi}_a   \,  ,
\end{equation}

\begin{equation}\label{sd2}
 \!\!\!\!\!\!\!\!  {1 \over 2} \partial_{\pm} \bphi^2  \,  =   \,
{i e   \kappa \bphi^2 \over \left[ \kappa^2 - e^2 \bphi^2
\right]^{1/2}  }\,
\hat{\psi}_a \bar{A}_{\pm}^a
  \,  ,
\end{equation}

where the  upper (lower) sign corresponds to positive (negative)
value of the
magnetic
flux.    \Eq{sd1}  implies  that the magnetic field vanishes whenever
$\bphi$
does.   The finiteness energy condition forces   the scalar field to
vanish
both at the center of  the vortex and also at spatial infinity;
consequently
as it happens in the Abelian case  \cite{torres},
the  magnetic flux of the vortices  lies in a ring.
It is interesting to remark that \Eq{sd2} can be written as a
self-duality
equation;
indeed if we define a new covariant derivative as
$\tilde{D}_i = \partial_i - i  2 e \kappa/\sqrt{\kappa^2 - e^2
\bphi^2}$,  then
\Eq{sd2}
is equivalent to

\begin{equation}\label{sd3}
   \tilde{D}_i   \bphi^2  \,  =   \,  \mp i   \epsilon_{ij}
\tilde{D}_j
\bphi^2
  \,  .
\end{equation}

     Eqs.  (\ref{sd1}) and   (\ref{sd2}) can be reduced
to one nonlinear second order differential  for one unknown function.
To do this, first notice that  \Eq{sd2}
implies  that  $\hat{\bpsi} \cdot  \bar{{\bf A}}_i$ can be determined
in terms
of the scalar field as

\begin{equation}\label{nrel}
\hat{\bpsi}  \cdot  \bar{{\bf A}}_i  = \pm  {1   \over  2 \kappa
\sqrt{C} }
\epsilon_{ij} \partial_j
\ln{\left(\bphi^2  \right) }
\, ; \end{equation}
when this equation is substituted into  \Eq{sd1}, we get

\begin{equation}\label{neq1}
  \partial_i \left[ { \sqrt{\kappa^2 - e^2 \bphi^2}  \over 2 e
\kappa}
\partial_i
 \ln{\left(\bphi^2  \right) }  \right]  \mp
 { 1 \over e}  \epsilon_{ij} \hat{\bpsi} \cdot   \left(\partial_i
\hat{\bphi}
 \wedge  \partial_j \hat{\bphi}  \right)  -
{\kappa e \bphi^2 \over \sqrt{\kappa^2 - e^2 \bphi^2}}  \, = \, 0
\, .
\end{equation}
This    equation still involves the three components of the scalar
field
$(\bphi)$,
but taking into account the gauge condition
$\epsilon_{ij} \hat{\bpsi} \cdot   \left(\partial_i \hat{\bphi}
\wedge
\partial_j \hat{\bphi}
\right )= 0$ and defining $ \sigma =  e^2 \bphi^2$ we  find that
\Eq{neq1} is a
second
order nonlinear differential equation for $\sigma$:

\begin{equation}\label{neq2}
  \partial_i \left[  \sqrt{\kappa^2 -  \sigma }   \partial_i
 \ln{\sigma   }  \right]   -
{2  \kappa^2 \sigma  \over \sqrt{\kappa^2 -  \sigma} }  \, = \, 0
\, . \end{equation}

If we assume a rotationally invariant form  for $ \bphi^2$, this
equation
reduces to the
Equation (16) of   reference \cite{torres} and therefore we  expect
that the
same features
of the Abelian vortex  solution  are present in  the non-Abelian
case.
However, we first have to   show that  it is possible to find a
vortex
configuration  that is consistent with the
conditions imposed on $\bpsi$  (\ref{cond2}) and the gauge fixing
conditions
(\ref{gauge}).
In this respect  the ansatz  of reference \cite{cuglian2}  fulfils
these
requirements; this axially
symmetric ansatz is given by

\begin{eqnarray}\label{ansatz}
\bphi  &= &  {\kappa \over e} f(\rho) \,  ( \cos{\theta},
\sin{\theta}, 0) \, ,
 \qquad
\bpsi = v \hat{e}_3 \, ,  \cr \cr
{\bf A}_\theta  &=&  \hat{e}_3 { a(\rho) - 1 \over e \rho }  \, ,
\qquad
\qquad \qquad
{\bf A}_0  =   \hat{e}_3  {\kappa \over e}  h(\rho)
\, ,
\end{eqnarray}
where $(\rho, \theta)$ are polar coordinates and $ \hat{e}_3=
(0,0,1)$ is a
unit
vector in isospin-space. It is straightforward to check that this
ansatz
satisfies the conditions
 (\ref{cond2})  and  (\ref{gauge}).  Substitution of this ansatz into
Eqs.
(\ref{sd1}) and  (\ref{sd2})
leads  to radial equations  that coincides with those found  for the
Abelian
case
 \cite{torres}

\begin{equation}\label{neq3}
{da \over d \rho}= {\pm} { \kappa^2 \rho  f^2 \over \left(1 - f^2
\right)^{1/2}}
\, ,  \qquad
{df \over d \rho} = {\mp} { f a \over \rho \left(1 - f^2
\right)^{1/2}}
\, .
\end{equation}

All the features of the non-topological vortex solution thus
corresponds,
for this  special ansatz, to those found in \cite{torres} if the
vorticity
number $n$
is chosen as $n=1$. In particular, the magnetic flux for these
solutions is
concentrated
in a ring surrounding the  center of the vortex.
The solution is characterized by a real valued constant $\alpha$,
that is
related with the asymptotic behavior of the fields.  Near the origin
the fields
behave as:
$f \to \rho$, $a \to 1$;  whereas at spatial infinity we have
$f \to \rho^{-\alpha} $, $a \to - \alpha$. The parameter
$\alpha$  satisfies the bounds $ 1 < \alpha < 3$ \cite{torres3}.
Once that  the boundary conditions are known  the magnetic flux can
be
calculated,  using the ansatz  \Eq{ansatz} and the definition of the
Abelian field strength \Eq{fmunu}, as
$\Phi_B = ( 2\pi/ e) \left( 1 + \alpha \right)$; notice that the
magnetic flux
is not quantized.
The soliton is also characterized by a charge, that can be directly
computed
from the fundamental relation \Eq{fundamental},
it results in
\begin{equation}\label{charge}
 Q  \, = \,     { 2\pi \kappa \over  e} \left( 1 + \alpha \right)
   \, = \,  { e \over 2}  \left( 1 + \alpha \right) \, n
\, ,
\end{equation}
where \Eq{cuank} was used  to write the last equality and
we recall that $n$ is an integer associated with the  quantization
of the CS term.  We can also compute
 the angular momentum (spin) of the vortex.  In $(2+1)$ dimensions
there
is only one generator of angular momentum
$ J = \int  d^2x \epsilon^{ij} x_i T_{0j}$;  using the expression
(\ref{tmunu}) for the energy-momentum tensor,  $J$ becomes

\begin{equation}\label{spin}
J  \, = \, {\pi \kappa \over e^2} \left( \alpha^2 -1  \right)
 \, = \, {1 \over 4}  \left( \alpha^2 -1  \right) \, n \, .
\end{equation}

Let us notice that
due  to  the non-topological nature of the solitons
the magnetic flux is not quantized.
However,  in the present non-Abelian model
both the  electric charge $Q$ and  the angular momentum $J$ are
quantized, a
condition that follows from   the
quantization of the CS  coefficient \Eq{cuank}.
Here, we conclude with some comments in relation
to the statistical parameter $\Delta \theta$.
According to the fundamental  CS relation \Eq{fundamental}
in $ 2 + 1$ dimensions, each charge becomes a flux tube, and
viceversa. Thus,
the effect of the CS term is  to transmute
the statistics of the particles. Indeed
  the statistical parameter  $\Delta \theta$,
arising  when  objects carrying magnetic flux and electric charge
winds
around another, is given by  \cite{wil}

\begin{equation}\label{sta}
\Delta \theta \, = \,   { 1\over 2} Q \Phi_B \, = \,
 { 1\over 2}   {Q^2 \over \kappa}\, = \,  { 1\over 2}
\kappa \Phi_B^2
   \, .
\end{equation}
This  relation leads to fractional statistics. For the present
non-topological
solitons and  using
Eqs.   (\ref{cuank}) and   (\ref{charge}) the statistical parameter
results in $\Delta \theta = (n  \pi /2)\left(1 + \alpha  \right)^2$.
Then, according to \Eq{spin} the statistical parameter is related
to the spin as

\begin{equation}\label{sta2}
\Delta \theta \, = \,  2 \pi {  \left( \alpha + 1 \right) \over
  \left( \alpha -1 \right) } J
\, ,
\end{equation}
instead of the usual spin-statistics  relation
$ |\Delta \theta | = 2 \pi J$. We  recall that $\alpha$ is related
to the asymptotic value of the gauge field; for topological solitons
$a(\rho)$
in \Eq{ansatz}   vanishes at  spatial infinity, so the
spin-statistics relation holds \cite{cuglian2}. Instead, for
non-topological
solitons the value $a(\infty) = - \alpha$ is not fixed  and leads to
the
modification of  the spin-statistics
relation  according to \Eq{sta2}.

\vskip 2.0cm

G.G. acknowledges partial support from CONACyT under grant 1199-E
9203.

\end{document}